\documentstyle[12pt]{article}
\begin{document}
\begin{flushright}
EPHOU-95-004\\
July, 1995 \\
\end{flushright}
\vspace{10mm}
\centerline{\LARGE Chiral Anomaly and Effective Field Theory}
\vspace{5mm}
\centerline{\LARGE for the Quantum Hall Liquid with Edges}
\vspace{10mm}
\centerline{\Large Nobuki Maeda}

\vspace{10mm}
\centerline{Department of Physics, Hokkaido University, Sapporo 060,
Japan}
\vspace{15mm}
\centerline{\Large abstract}
\baselineskip=24pt
Under general assumptions, we present a low-energy effective action
for the quantum Hall state when edges exist.
It is shown that the chiral edge current is necessary to make the
effective action to be gauge invariant.
However the chiral edge current is irrelevant to the Hall current.
The exactly quantized value of $\sigma_{xy}$ is observed only when
the Hall current does not flow at the edge region.
Our effective theory is applicable to the quantum Hall liquid on
a surface with non-trivial topology and physical meanings of the
topology are discussed.

\newpage
\baselineskip=24pt

There are two approaches to the quantum Hall effect(QHE)\cite{a}.
In the one approach, the electric current is carried by bulk
states and the boundary condition is not important.
The Hall conductance is represented as a topological invariant and
is quantized exactly\cite{d}.
In the other approach, the electric current is carried by edge states
and the bulk effects are neglected.
The Hall current is proportional to the difference of
chemical potentials at the edges and the Hall conductance is
proportional to a number of edge current modes\cite{g}.
Recently the relation between the bulk picture and the edge picture is
discussed by many authors\cite{i}.
In this letter, we analyze a low-energy effective field theory
which is applicable to the quantum Hall liquid(QHL)
with edges, no matter where the Hall current flows.
It will be shown that the chiral edge current is canceled in the
total Hall current.

We consider that two-dimensional
spinless electrons are in a perpendicular
uniform magnetic field $B$ and use the unit
$\hbar=c=1$.
For a while, we suppose that electrons are bounded in negative
$x_2$-direction by an electrostatic potential $V(x)$.
Due to $V(x)$, a diamagnetic edge current flows at the edge.
We denote this current density as $J^\mu_{(0)}$ and
$\partial_\mu J^\mu_{(0)}=0$.
The external field $A_\mu$ couples to electrons
in a gauge invariant manner in addition to
$B$ and $V(x)$.
The effective action for $A_\mu$ is obtained by integrating out the
electron field.
The following action should be generated by electrons in the bulk
states,
\begin{equation}
S_{\rm CS}={\sigma_{xy}\over 2}\int d^3 x f_L(x_2)\epsilon^{\mu\nu\rho}
A_\mu\partial_\nu A_\rho.
\end{equation}
If $f_L$ is constant, this is called the Chern-Simons action\cite{b}.
This action results from a parity violation in the presence
of a magnetic field.
$f_L$ varies only in the edge region, $\vert x_2\vert<\delta,\
\delta=O((eB)^{-{1\over2}})$, and takes constant value in other regions.
We fix the value as
\begin{equation}
f_L(x_2)=\cases{1,&for $x_2<-\delta$,\cr 0,&for $x_2>\delta$.\cr}
\end{equation}
When electrons in bulk states are in the quantum Hall regime,
the Hall conductance is quantized as
\begin{equation}
\sigma_{xy}={e^2\over 2\pi}N,
\end{equation}
where $N$ is integer in the integer quantum Hall effect(IQHE) and
is rational number in the fractional quantum Hall effect(FQHE).
The electric current density is obtained from Eq.(1) as
\begin{equation}
J^\mu_{\rm CS}=-{\delta S_{\rm CS}\over \delta A_\mu}=
-\sigma_{xy}f_L \epsilon^{\mu\nu\rho}\partial_\nu A_\rho+
{\sigma_{xy}\over 2}\partial_2 f_L \epsilon^{2\mu\nu}A_\nu
\end{equation}
The second term of Eq.(4) is a edge current induced from
$S_{\rm CS}$.

The above argument is insufficient because the action $S_{\rm CS}$
is not gauge invariant in the edge region\cite{o}.
Under a gauge transformation, $A_\mu\rightarrow A_\mu+\partial_\mu
\phi$, the action $S_{\rm CS}$ is changed by
\begin{equation}
\Delta S_{\rm CS}=-{\sigma_{xy}\over 2}
\int d^3x\partial_2 f_L \epsilon^{2\mu\nu}\partial_\mu A_\nu
\phi,
\end{equation}
where an integration of total derivative is omitted.
Moreover the current density $J^\mu_{\rm CS}$ has an anomaly,
\begin{equation}
\partial_\mu J^\mu_{\rm CS}=-\sigma_{xy}\partial_2 f_L
\epsilon^{2\mu\nu}\partial_\mu A_\nu.
\end{equation}
The anomaly comes from the anomalous current in Eq.(4).
Since the anomalous current is quasi-one-dimensional, the anomaly
can be canceled by a quasi-one-dimensional effective action.
It is noted that the quasi-one-dimensional effective action can not be
obtained by $A_\mu$-integration of anomalous current,
because the anomalous current is non-integrable.

{}From a physical point of view, it is natural to expect that the
necessary quasi-one-dimensional effective action should be generated
from electrons in edge states.
At the edge region, gapless electron states are exist and these
states are extended in $x_1$-directoin.
Since the edge current flows in one direction, electrons in edge states
can be described by the chiral fermion and the action is
\begin{equation}
S_{\rm edge}(\Psi,A)=\int d^3 x(\bar\Psi i\gamma^i\partial_i\Psi-
{1\over2}g_L(x_2)\sqrt{\Delta x_2}\bar\Psi\gamma^i(1-\gamma_5)\Psi
A_i),
\end{equation}
where $i=0,1$ and $\gamma^i$'s are Dirac matrices in $1+1$ dimensional
spacetime.
In Eq.(7), we rescale $x_1$-component of vectors by a Fermi
velocity $v_F$ as $v_F\partial_1\rightarrow\partial_1$ and
$v_F A_1\rightarrow A_1$.
Hereafter we omit a Fermi velocity $v_F$ for convenience,
because our arguments are not affected by this rescaling
essentially.
Only the left handed fermion couples to the gauge potential $A_i$.
In Eq.(7), $\Delta x_2$ is a short distance cutoff.
This cutoff is necessary to renormalize a divergence in a current-current
correlation function.
Since the kinetic term does not include $x_2$-derivative,
one-loop diagram diverges as $\int dp_2\sim\Lambda_2\sim 1/\Delta x_2$.
Current-current correlation function includes two vertices and its
one-loop diagram behaves as (coupling)$^2\sim\Delta x_2$.
Then, we obtain a finite effective action as $\Delta x_2\rightarrow0$.
$g_L$ is a coordinate-dependent coupling and is determined by
the gauge invariance later.
The quasi-one-dimensional effective action and chiral edge current
density are obtained from Eq.(7) as
$$
e^{iS_{\rm edge}(A)}=\int D\Psi e^{iS_{\rm edge}(\Psi,A)},
$$
$$
S_{\rm edge}(A)={1\over 8\pi}\int d^3x g^2_L(x_2)
A_i[\alpha g^{il}-
(g^{ij}+\epsilon^{ij}){\partial_j\partial_k\over\Box}
(g^{kl}-\epsilon^{kl})]A_l,
$$
\begin{equation}
J^i_L=-{g^2_L(x_2)\over 4\pi}[\alpha g^{il}-(g^{ij}+
\epsilon^{ij}){\partial_j\partial_k\over\Box}
(g^{kl}-\epsilon^{kl})]A_l,
\end{equation}
where $\Box=\partial^i\partial_i$ and $\alpha$ is regularization
dependent parameter\cite{l}.
The current density $J^i_L$ has an anomaly
\begin{equation}
\partial_i J^i_L={g^2_L(x_2)\over 4\pi}[(1-\alpha)\partial_i A^i-
\epsilon^{ij}\partial_i A_j].
\end{equation}
{}From Eqs.(6) and (9), we obtain an anomaly free current density,
\begin{equation}
J^\mu_{\rm Total}=J^\mu_{\rm CS}+J^\mu_L+J^\mu_{(0)},
\end{equation}
$$
\partial_\mu J^\mu_{\rm Total}=0,
$$
where $J^\mu_{(0)}$ is a diamagnetic current due to $V(x)$,
by setting
\begin{equation}
\alpha=1,\quad g^2_L(x_2)=-2\pi\sigma_{xy}\partial_2 f(x_2).
\end{equation}
In the quantum Hall regime, the coupling $g_L$ is quantized as
\begin{equation}
\int^\infty_{-\infty} g^2_L(x_2)dx_2=Ne^2,
\end{equation}
using Eqs.(2) and (3).
Furthermore the total effective action,
\begin{equation}
S_{\rm Total}=S_{\rm CS}+S_{\rm edge}+\int J^\mu_{(0)}A_\mu d^3x,
\end{equation}
becomes gauge invariant if and only if Eq.(11) is satisfied.

This anomaly cancelation was already discussed by X. G. Wen
in Ref.\cite{o}.
He studied the QHL on the plane with boundary
and showed the existence of the edge currents satisfying the
U(1) Kac-Moody algebra.
Our theory is generalization of Wen's theory to the case that
the edge region has a finite width.
We study not only
the chiral edge current but also the current induced from the Chern-Simons
action $S_{\rm CS}$ and show that the edge current is irrelevant
to the Hall conductance.

Next we consider the case that two edges exist.
Electrons are bounded in the region
$-L<x_2<0$ by an electrostatic potential $V(x)$.
In this case a function $f_L(x_2)$ is replaced with $f(x_2)$ which is
given as
\begin{equation}
f(x_2)=\cases{f_L(x_2),&for $x_2>-L/2$,\cr
f_R(x_2),&for $x_2<-L/2$,\cr}
\end{equation}
$$
{\rm where\qquad\qquad }
f_R(x_2)=\cases{1,&for $x_2>-L+\delta$,\cr
0,&for $x_2<-L-\delta$.\cr}
$$
Following the previous argument of gauge invariance of $S_{\rm CS}$,
an effective action for the chiral fermion is required as
\begin{equation}
S_{\rm edge}(\Psi,A)=\int d^3 x(\bar\Psi i\gamma^i\partial_i\Psi-
{1\over2}g_L(x_2)\sqrt{\Delta x_2}\bar\Psi\gamma^i(1-\gamma_5)\Psi
A_i)
\end{equation}
$$
\qquad
-{1\over2}g_R(x_2)\sqrt{\Delta x_2}\bar\Psi\gamma^i(1+\gamma_5)\Psi
A_i),
$$
$$
{\rm where\qquad\qquad }
g^2_R(x_2)=2\pi\sigma_{xy}\partial_2 f_R(x_2).
$$

Now we consider the system with two edges in the electric field
$E_2(x_2)$ and use the following gauge choice,
\begin{equation}
E_2(x_2)=-\partial_2 A_0(x_2),
\end{equation}
$$
A_1=A_2=0.
$$
By substituting Eq.(16) into Eqs.(4) and (8) with Eq.(11),
$J^1_{\rm CS}$ and $J^1_L+J^1_R$ are calculated as
\begin{equation}
J^1_{\rm CS}=-\sigma_{xy}f(x_2)\partial_2 A_0-
{\sigma_{xy}\over 2}\partial_2 f(x_2) A_0,
\end{equation}
$$
J^1_L+J^1_R={\sigma_{xy}\over 2}\partial_2 f(x_2) A_0.
$$
Thus chiral edge current densities $J^1_L$ and $J^1_R$ are canceled
by the edge current in $J^1_{\rm CS}$ and total current density
becomes
\begin{equation}
J^1_{\rm Total}=-\sigma_{xy}f(x_2)\partial_2 A_0+J^\mu_{(0)}.
\end{equation}
This means that the chiral edge current is irrelevant in the Hall
current.
{}From Eq.(18) the distribution of the Hall current depends on
$f(x_2)$ and $E_2(x_2)$ and the total current $I^1$ is given by
\begin{equation}
I^1=\int^\infty_{-\infty}J^1_{\rm Total}dx_2=\sigma_{xy}(
\langle A_0\rangle_L-\langle A_0\rangle_R),
\end{equation}
$$
\langle A_0\rangle_{L}=-\int \partial_2 f_{L}(x_2)A_0 dx_2,
$$
$$
\langle A_0\rangle_{R}=+\int \partial_2 f_{R}(x_2)A_0 dx_2,
$$
where we assumed that $V(x)$ is so choosen that the diamagnetic current
vanishes, that is $I^1_{(0)}=\int J^1_{(0)}dx_2=0$.
If $f(x_2)$ is a step-like function then $\vert\partial_2 f(x_2)
\vert$ is well
localized normalized function in $x_2$-direction.
Therefore $\langle A_0\rangle_{\rm edge}$ is the average value of the
potential at the edge region.
The relation of Eq.(19) is independent of the distribution of
$J^1_{\rm Total}(x_2)$.

The current distribution is determined by the electric field $E_2$
in Eq.(18).
Since the electric field depends on the detail of electron density
and experimental settings, the current distribution cannot be
calculated by the low-energy effective action $S_{\rm Total}$.
What we can say from Eq.(19) is that the exactly quantized value
of $\sigma_{xy}$ is observed as a ratio of total current and total
voltage only when the potential $A_0$ is constant and
the Hall current does not flows at the edge region.
The quantized value of $\sigma_{xy}$ is observed approximately when
the potential $A_0$ can be regarded as constant compared with
a magnetic length in the edge region.

If the electrostatic potential $V(x)$ is very smooth, then
$f(x)$ becomes steps-like function and many edges appear.
See Fig.1.
For IQHE, each step has a height $1/N$ and a number of steps is $N$.
We denote the position of edges as $R_i$ and $L_i$ $(i=1\sim N)$
at right and left sides of QHL.
We assume that each edge has a width $\delta$.
Then $f(x_2)$ satisfies,
\begin{equation}
-\int^{L_i+\delta}_{L_i-\delta}\partial_2 f dx_2={1\over N},
\end{equation}
$$
-\int^{R_i+\delta}_{R_i-\delta}\partial_2 f dx_2=-{1\over N}.
$$
For the gauge invariance, N-chiral edge currents are needed and the
relation (19) is changed as,
\begin{equation}
I^1=\sigma_{xy}\sum^N_{i=1}(\langle A_0\rangle_{L_i}-
\langle A_0\rangle_{R_i})/N,
\end{equation}
$$
\langle A_0\rangle_{L_i}=-N\int^{L_i+\delta}_{L_i-\delta}
\partial_2 f(x_2)A_0 dx_2,
$$
$$
\langle A_0\rangle_{R_i}=+N\int^{R_i+\delta}_{R_i-\delta}
\partial_2 f(x_2)A_0 dx_2.
$$
If each side of the QHL is in equilibrium, then
$\langle A_0\rangle_{L_1}=\langle A_0\rangle_{L_2}=\cdots=
\langle A_0\rangle_{L_N}$ and
$\langle A_0\rangle_{R_1}=\langle A_0\rangle_{R_2}=\cdots=
\langle A_0\rangle_{R_N}$ and Eq.(21) is equivalent to Eq.(19).
But if the left side of the QHL is not in equilibrium and
$\langle A_0\rangle$'s satisfy,
\begin{equation}
\langle A_0\rangle_{R_1}=\langle A_0\rangle_{R_2}=\cdots=
\langle A_0\rangle_{R_N}=\langle A_0\rangle_R,
\end{equation}
$$
\langle A_0\rangle_{L_1}=\langle A_0\rangle_{L_2}=\cdots=
\langle A_0\rangle_{L_K}=\langle A_0\rangle_R,
$$
$$
\langle A_0\rangle_{L_{K+1}}=\langle A_0\rangle_{L_{K+2}}=\cdots=
\langle A_0\rangle_{L_N}=\langle A_0\rangle_L,
$$
then Eq.(21) becomes,
\begin{equation}
I^1={e^2\over 2\pi}(N-K)(\langle A_0\rangle_L-\langle A_0\rangle_R).
\end{equation}
This anomalous IQHE was observed experimentally\cite{r}.
Same relation of Eq.(23) is also derived by using the
B\"uttiker-Landauer formula in the edge picture.
Our picture is more general than
the edge picture because even if the Hall current
flows only
in the region between two edges in the left side region,
Eq.(23) is valid.

It is interesting to consider the case that the electric field is
parallel to the edges.
That is
\begin{equation}
E_1(x_1)=-\partial_1 A_0(x_1),
\end{equation}
$$
A_1=A_2=0.
$$
We consider a single-step edge of Eq.(2) here.
The edge current density
in $J^\mu_{\rm CS}$ is the second term in Eq.(4),
and the edge current $I^i_{\rm CS,edge}$ and $I^i_L$ are given by
\begin{equation}
I^i_{\rm CS,edge}=\int J^i_{\rm CS,edge}dx_2=
{\sigma_{xy}\over 2}\int \partial_2 f\epsilon^{ij}A_j dx_2=
-{\sigma_{xy}\over 2}\epsilon^{ij}A_j,
\end{equation}
$$
I^i_L=\int J^i_L dx_2,\quad(i=0,1).
$$
Using Eqs.(24), (25), and (9) with Eq.(11), we obtain
\begin{equation}
\partial_i I^i_{\rm CS,edge}=-{\sigma_{xy}\over 2}E_1,
\end{equation}
$$
\partial_i I^i_L=-{\sigma_{xy}\over 2}E_1.
$$
Thus the total edge current,
$I^i_{\rm Total,edge}=I^i_{\rm CS,edge}+I^i_L$, satisfies
\begin{equation}
\partial_i I^i_{\rm Total,edge}=-\sigma_{xy}E_1.
\end{equation}
We can easily show that $I^1_{\rm Total,edge}$ vanishes.
Then Eq.(27) means that the current flowing perpendicular to the edge
equals to $-\sigma_{xy}E_1$ and the QHE occurs.

In the case of Eq.(16), it is important that anomalies of $J_{\rm CS}$
and $J_L$ cancel.
On the other hand, in the case of Eq.(24), sum of each anomaly
contributes to the Hall current in Eq.(27).

Using Eq.(27), we can calculate the charge of a quasiparticle moving
on edges in the excited states\cite{o}.
Excited states can be constructed by adding a fixtitious flux
$\phi_0=2\pi/e$ to the bulk states.
Using the Bianchi identity, $\epsilon^{\mu\nu\rho}\partial_\mu F_{\nu
\rho}=0$, a charge of quasiparticle is given by
\begin{equation}
Q=\int dtdx_1\partial_0 I^0_{\rm Total,edge}=
-\int dtdx_1 \sigma_{xy}E_1=\sigma_{xy}\int dt \partial_0\phi=
\sigma_{xy}\phi_0=Ne.
\end{equation}
In FQHE, $N$ takes fractional value and the
quasiparticle has a fractional charge.
In order
for the system to be not changed by insertion of flux $\phi_0$ at $x_0$,
$f(x)$ in $S_{\rm CS}$ has to vanish at $x_0$.
This corresponds to the quasiparticle in the bulk states\cite{n}
which has a charge $-Ne$.
These excitations would have complex dynamics and can not be described
by our low-energy effective field theory.
To describe these excitations, we have to treat $f(x)$ in $S_{\rm CS}$
as a dynamical quantity.
Then we should regarded $f(x)$ as an order parameter for topological
states of QHL.

We can imagine a two-dimensional phase diagram for $f$ and
$\partial f$.
A line of $f=0$ is a parity unbroken phase and
a line of $\partial f=0$ is a topological phase.
There are stable points on the topological phase line.
These points correspond to QHL states.
The QHL with edges is represented as a path linking these stable
points.
The path is parametrized by $x_2$, $-\infty<x_2<\infty$.
Fig.2 shows a single-step case($P_1$) and a double-step case($P_2$)
for N=2 IQHE.

Finaly we generalize previous arguments to the QHL
on a surface with non-trivial topology.
That is a surface with boundaries, punctures, and handles.
For simplicity we take $\delta=0$ and edges of QHL
can be regarded as boundaries of two-dimensional surface.
Since the action $S_{\rm CS}$ is diffeomorphism invariant without
metric of the surface(topological invariant),
we can deform the surface as boundaris to become straight lines.
Thus previous arguments on the edge are valid in the case of curved
edges.
A handle is regarded as a cylinder attached to two boundaries
of the surface and previous arguments of two edges are also
applicable to it.
Each handle has a pair of non-contractable loops $l_1,l_2$.
Without changing a magnetic field on the surface, fluxes $\phi_1,\phi_2$
can go through the loops.
The flux and the current flowing across the loop $l_i$ are related
as
\begin{equation}
I_i=\sigma_{xy}\int_{l_i}\epsilon^{ij}E_j=\sigma_{xy}\dot\phi_i,
\quad (i=1,2).
\end{equation}
Thus a handle works as a current source and a current drain.
A puncture is an infinitesimal hole of the surface and is represented
as a point where $f(x)$ vanishes in our theory.
Without changing a magnetic field on the surface, unit flux can be
inserted into the puncture.
Then a puncture is geometric representation of a quasiparticle in the
bulk states.

In conclusion, we obtained a low-energy effective theory for the
QHL with edges.
Although chiral edge currents are necessary for the gauge invariance
of the effective action, it does not contribute to the Hall current.
It is shown that the quantization of $\sigma_{xy}$ is observed exactly
only when the Hall current does not flow in the edge regions.
The anomalous IQHE was explained in our theory.
Using a function $f$ as an order parameter,
we presented a phase diagram for the QHL with edges.
Physical meanings of topology of the surface where electrons live
were clarified.

I am very grateful to Professor K. Ishikawa for his suggestion to
consider this subject and for useful discussions.

\newpage
\centerline{\Large Figure Caption}

Fig.1 :
$f(x)$ with two steps for the N=2 IQHE.

Fig.2 :
A phase diagram for the N=2 QHL with one step($P_1$) and
two steps($P_2$).
The paths in the upper half plane represent right side edges of
the QHL and paths in the lower half plane represent left side edges
of the QHL.

\end{document}